\input{aipcheck.tex}
\documentclass[article]{aipproc}
\layoutstyle{6x9}
\begin{document}

\title{Form factors of baryons within the framework of light-cone sum rules}


\author{Alexander J. Lenz}{
  address={Fakult{\"a}t f{\"u}r Physik, Universit{\"a}t Regensburg, 
           93040 Regensburg, Germany},
  email={alexander.lenz@physik.uni-regensburg.de}}

\begin{abstract}
We present the application of the method of light-cone sum rules
to the determination of baryonic form factors at intermediate momentum
transfer. After reviewing the current 
status of this field we give some outlook on possible future projects.
\end{abstract}

\date{\today}

\maketitle

\section{Introduction}
In this talk we present the current status of the
determination of baryonic form factors and of transition form factors within QCD.
Form factors are interesting quantities per se, since they encode information
about the structure of the investigated baryon.
This interest raised a lot in recent years, in particular because new data from JLAB 
\cite{JLAB1,JLAB2,JLAB3,JLAB4} for the well-known electromagnetic form factors of the nucleon
contradict common textbook-wisdom! 
See \cite{FFreview} for a review and references therein.
\\
In our approach - light-cone sum rules - we relate the form factors directly to the distribution
amplitude of a baryon, mostly the nucleon. Therefore one can follow two different philosophies:
\begin{itemize}
\item Start with a non-perturbative model (lattice, sum rules,...) for the distribution amplitude
      and determine the physical interesting  form factors.
\item Determine the non-perturbative distribution amplitude by fitting the experimental numbers
      of the form factors to the light-cone prediction.
\end{itemize}
In the following we will present the current status of these investigations.
\section{The Nucleon Distribution Amplitude}
The nucleon distribution amplitude has been determined up to contributions
of twist 6 in \cite{BFMS2000}. 
$x^2$-corrections to the leading twist distribution 
amplitudes were determined in \cite{BLMS2001,HW2004,BLW2006}.
In \cite{BLW2006} there was unfortunately a sign error in the contribution 
of $A_1^{M(u)}$, the corrected plots are presented below and 
the corrected formulas can be found in the appendix.
\\
Including next-to-leading terms in the conformal expansion the whole 
distribution amplitude is expressed in terms of eight non-perturbative 
parameters. One can write
\begin{equation}
4 \langle 0 |\epsilon_{ijk} 
u_{\alpha}^i (a_1 x)
u_{\beta}^j  (a_2 x)
d_{\gamma}^k (a_3 x) | N(P)\rangle
= \sum \Gamma_3^{\alpha \beta} \Gamma_4^{\gamma} F \, ,
\end{equation}
where $\Gamma_{3/4}$ are certain Dirac structures, $N$ describes the nucleon
state, $a_i$ are positive numbers with $a_1+a_2+a_3=1$ and the $F$ are 
distribution amplitudes depending on the non-perturbative parameters
$f_N$, $\lambda_1$, $\lambda_2$, $V_1^d$, $A_1^u$, $f_1^d$, $f_1^u$ 
and $f_2^d$, for details see \cite{BFMS2000,BLW2006}.
As in the meson case these parameters can be estimated with QCD sum rules 
\cite{QCDSR}
see e.g. \cite{BL2004,BBL2006,BBL2007,Thorsten} for some state of the art work in 
the meson case.
QCD sum rule estimates for all eight parameters of the nucleon
distribution amplitude were first presented in
\cite{BLMS2001} and later on updated in \cite{BLW2006}.
The latter parameter set will be called {\it sum-rule} estimate in the following.
Demanding that the next-to-leading conformal contributions vanish, fixes 
five of the eight parameters. This parameter set will be called 
{\it asymptotic}.
In \cite{BLW2006} we presented a third parameter set, called {\it BLW}:
with the help of light-cone sum rules \cite{LCSR1,LCSR2}
one can express the nucleon form factors
in terms of the eight non-perturbative parameters. Choosing values in between
the {\it asymptotic} and {\it sum-rule} ones, we got an astonishingly good 
agreement with the experimental numbers, see \cite{BLW2006}.
This procedure however does not replace the necessity of performing a real fit
after $\alpha_s$-corrections have been calculated to the light-cone sum rules.
Finally a lattice determination of these eight non-perturbative
parameters would be very desirable.
\section{Light-cone sum rules for form factors}
The starting point for our analysis is a correlation function of the 
following form.
\begin{equation}
T (P,q) = \int d^4x e^{-ipx} 
\langle 0 | T \{ \eta(0) j(x) \} | N(P) \rangle \, ,
\end{equation}
which describes the transition of a baryon $B(P-q)$ to the nucleon $N(P)$
via the current $j$. The baryon $B$ is created  by the interpolating 
three-quark field $\eta$, e.g. the Ioffe-current for the nucleon
\begin{equation}
  \eta_{\rm Ioffe}(x) = 
\epsilon^{ijk} \left[u^i(x) (C \gamma_\nu)\, u^j(x)\right] \, 
(\gamma_5 \gamma^\nu)\, 
d^k_{\delta}(x) \, .
\end{equation}
A typical example for $j$ is the electromagnetic current in the case of the
electromagnetic form factors
\begin{equation} 
j_{\mu}^{\rm em}(x) = e_u \bar{u}(x) \gamma_{\mu} u(x) + 
                       e_d \bar{d}(x) \gamma_{\mu} d(x) \, .
\end{equation}
The basic idea of the light-cone sum rule approach is to calculate this
correlation function both on the hadron level (expressed in terms of 
form factors) and on the quark level (expressed in terms of the nucleon
distribution amplitude). Equating both results and performing a Borel 
transformation to suppress higher mass states one can express the
form factors in terms of the eight non-perturbative parameters of the 
nucleon distribution amplitude and in terms of the Borel parameter $M_B$ 
and the continuum threshold $s_0$, for details see \cite{BLMS2001,BLW2006}.
\\ 
We studied the electromagnetic nucleon form factors with the
Chernyak-Zhitnitsky interpolating field in \cite{BLMS2001}.
In \cite{LWS2003} we found that $\eta_{CZ}$ yields to large unphysical isospin
violating effects, therefore we introduced a new isospin respecting CZ-like current to
determine the electromagnetic form factors.
In \cite{BLW2006} we also studied the Ioffe current for the nucleon and 
extended our studies from the electromagnetic form factors to axial
form factors, pseudoscalar form factors and the neutron to proton transition.
It turned out that the Ioffe current yields the most reliable results.
The $N \to \Delta$-transition was studied in this framework in
\cite{BLPR2005} (for a similar approach for $Q^2 = 0$ see e.g. \cite{R2007}),
while in \cite{BILP2006} we investigated pion-electroproduction.
We present the numerical results of these calculations in the next section.
\\
Moreover one can find the decay $\Lambda_b \to p l \nu$ within this approach
in \cite{HW2004}. 
Using $\eta_{CZ}$ the scalar form factor of the nucleon 
was considered in \cite{WW2006} and the axial and the pseudoscalar
one in \cite{WW2006b}.
The authors of \cite{HW2006} considered  $\Lambda_c \to \Lambda l \nu$
and therefore determined a part of the $\Lambda$ distribution amplitude.
In \cite{W2006} the transition $\Sigma \to N$ was determined.
Using a general form of the interpolating nucleon field the scalar form 
factor of the nucleon was considered again in \cite{AS2006}. Just recently
the axial part of the  $N \to \Delta$-transition was calculated in
\cite{AAO2007}.


\section{Numerical Results}

\begin{figure*}
  \includegraphics[width=0.45\textwidth,angle=0]{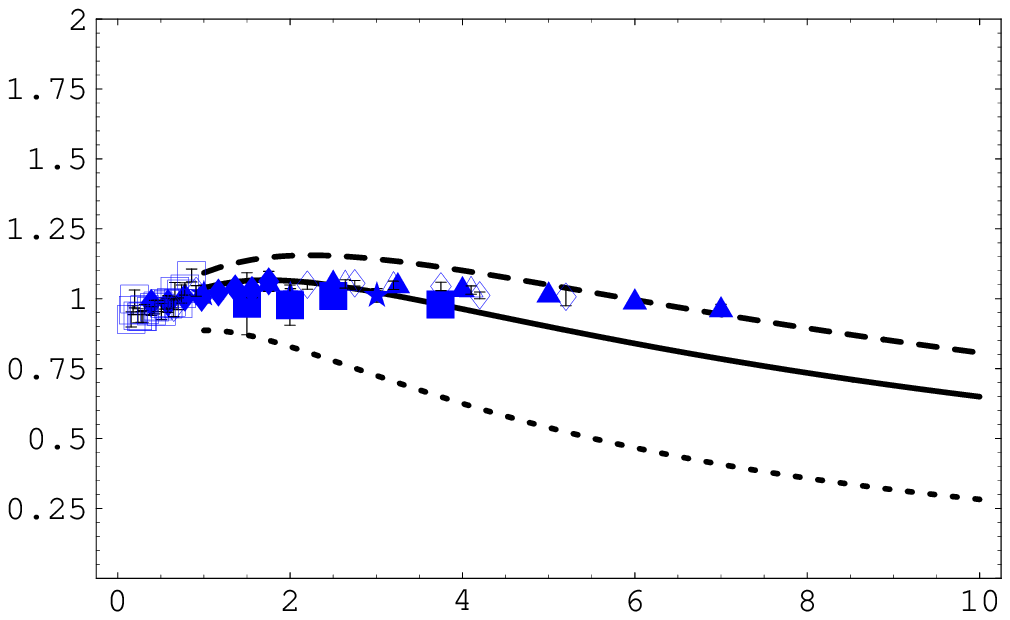}
  \includegraphics[width=0.45\textwidth,angle=0]{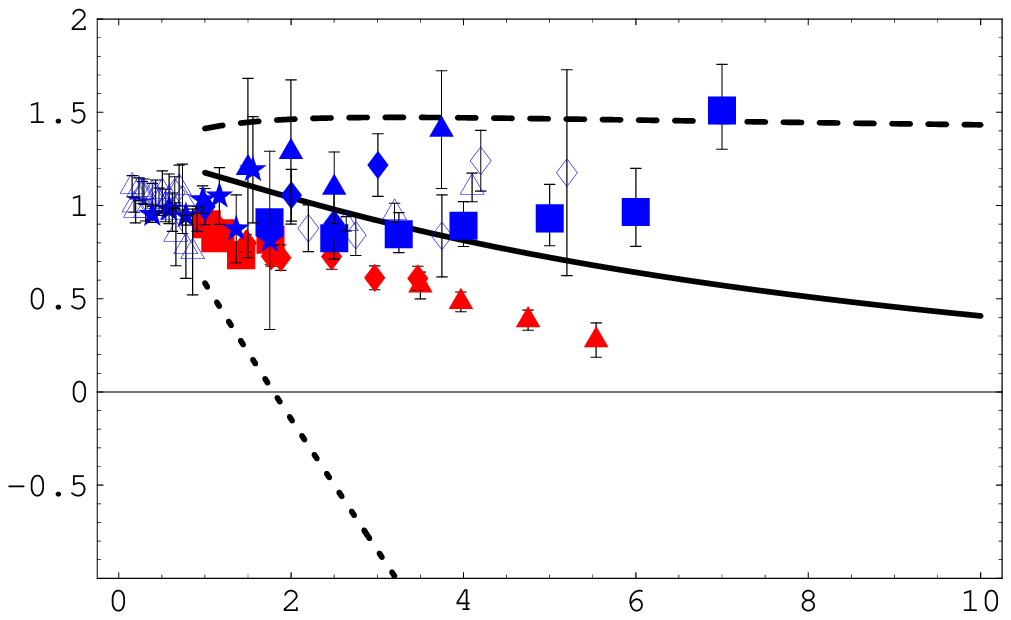}
\caption{LCSR results for the electromagnetic form factors 
(left: $G_M/( \mu_p G_{\it Dipole})$ vs. $Q^2$; right:  $ \mu_p G_E/G_M$ vs. $Q^2$) of the 
proton, obtained using the BLW model (solid line), 
the asymptotic model (dashed line)
and sum rule model (dotted line) of the nucleon DAs. 
The red data points on the right picture are JLAB data, while the blue ones
are obtained via Rosenbluth separation.
For the references to the experimental data see ~\cite{BLW2006}.
}
\label{fig:GMGEp}
\end{figure*}
\begin{figure*}
  \includegraphics[width=0.45\textwidth,angle=0]{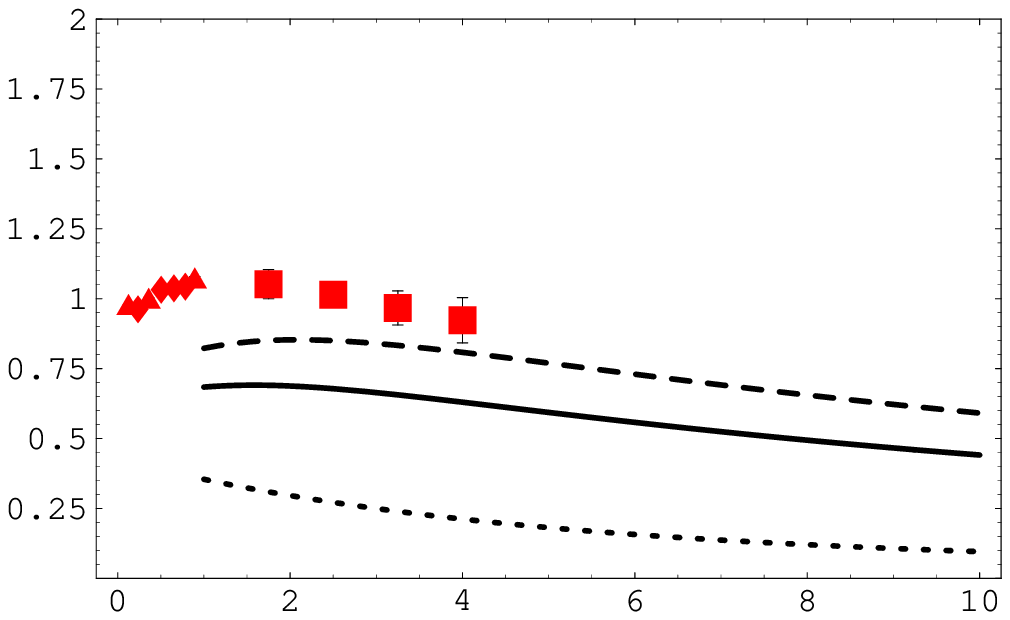}
  \includegraphics[width=0.45\textwidth,angle=0]{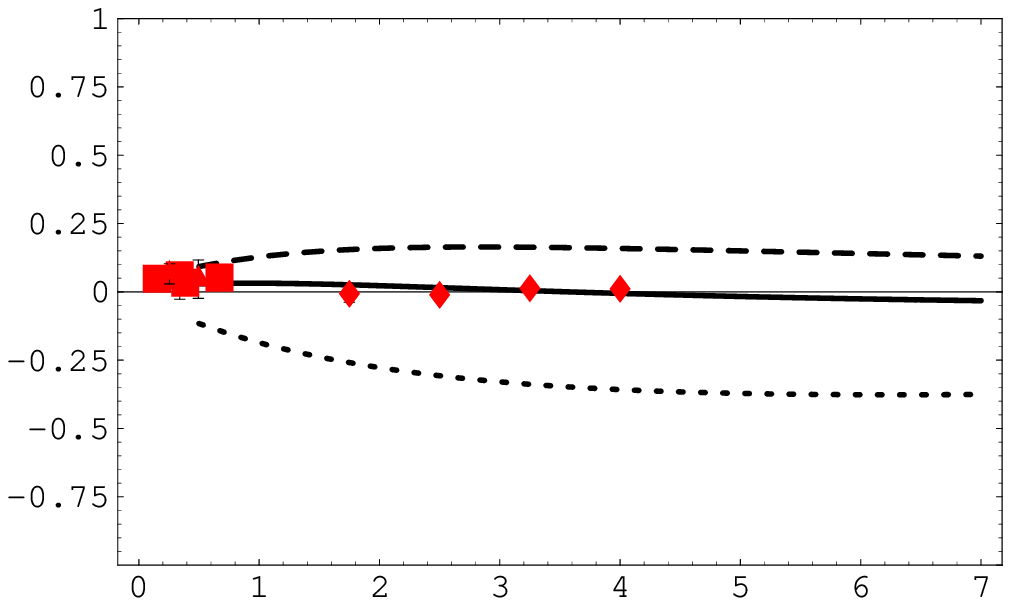}
\caption{LCSR results for the electromagnetic form factors of the 
neutron (left: $G_M/(\mu_n G_{\it Dipole})$ vs. $Q^2$; right:  $G_E$ vs. $Q^2$), 
obtained using the BLW model (solid line), 
the asymptotic model (dashed line)
and sum rule model (dotted line) of the nucleon DAs. 
For the references to the experimental data see ~\cite{BLW2006}.
}
\label{fig:GMGEn}
\end{figure*}
\begin{figure*}
  \includegraphics[width=0.44\textwidth,angle=0]{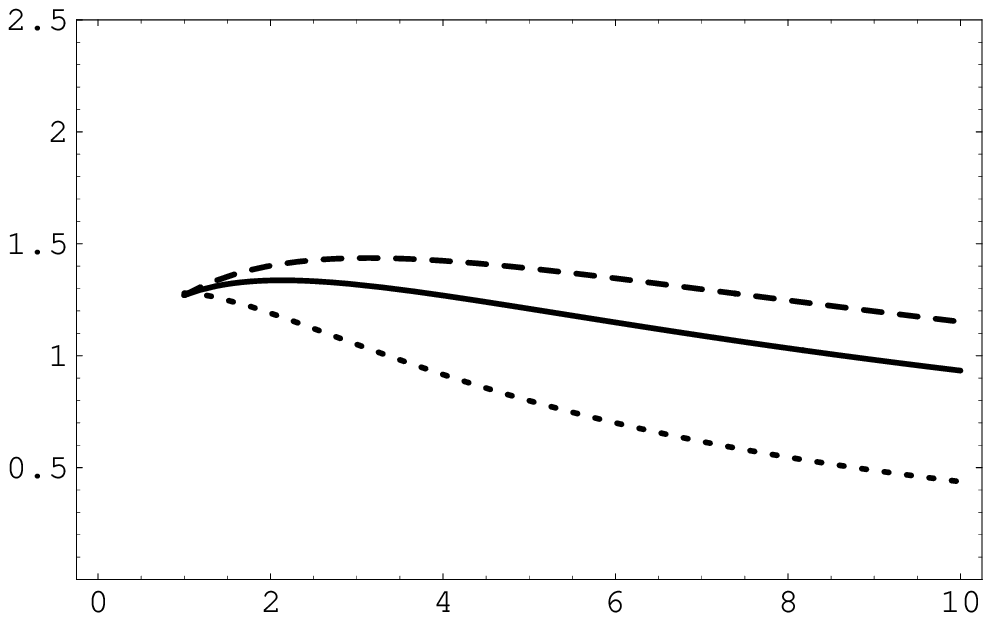}
  \includegraphics[width=0.45\textwidth,angle=0]{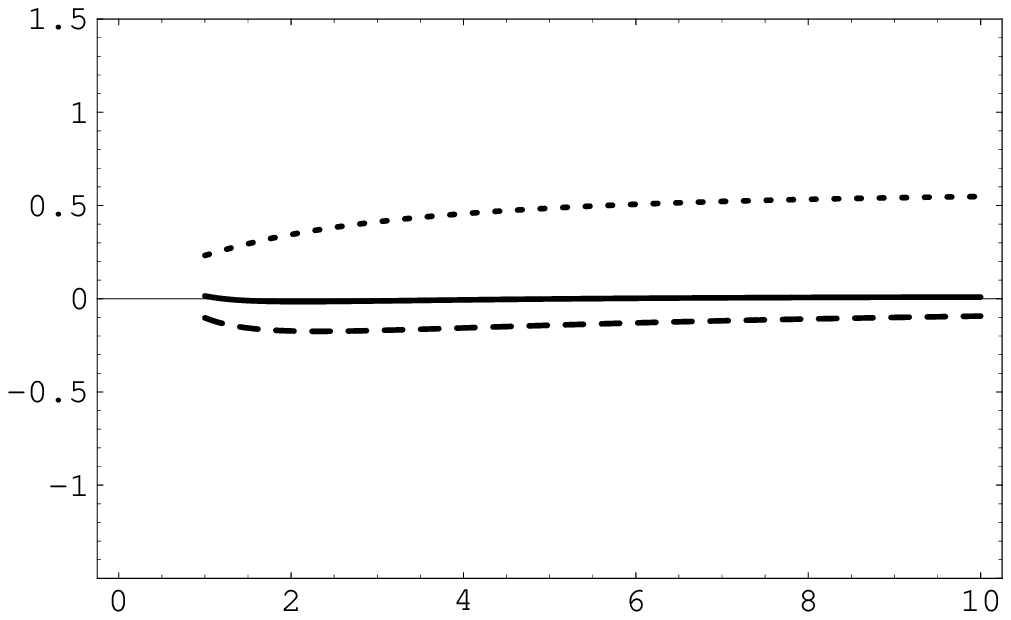}
\caption{LCSR results (solid curves) for the axial form factor  of the proton
$G_A^{CC}$ normalized to $G_D^{(a)}=g_A/(1+Q^2)^2$  vs. $Q^2$(left panel) and 
the tensor form factor $G_T^{CC}$ normalized to $G_A^{CC}$  vs. $Q^2$ (right panel),
obtained using the BLW model (solid line), 
the asymptotic model (dashed line)
and sum rule model (dotted line) of the nucleon DAs. 
}
\label{fig:GAGT}
\end{figure*} 

\begin{figure*}[ht]
  \includegraphics[width=0.33\textwidth,angle=0]{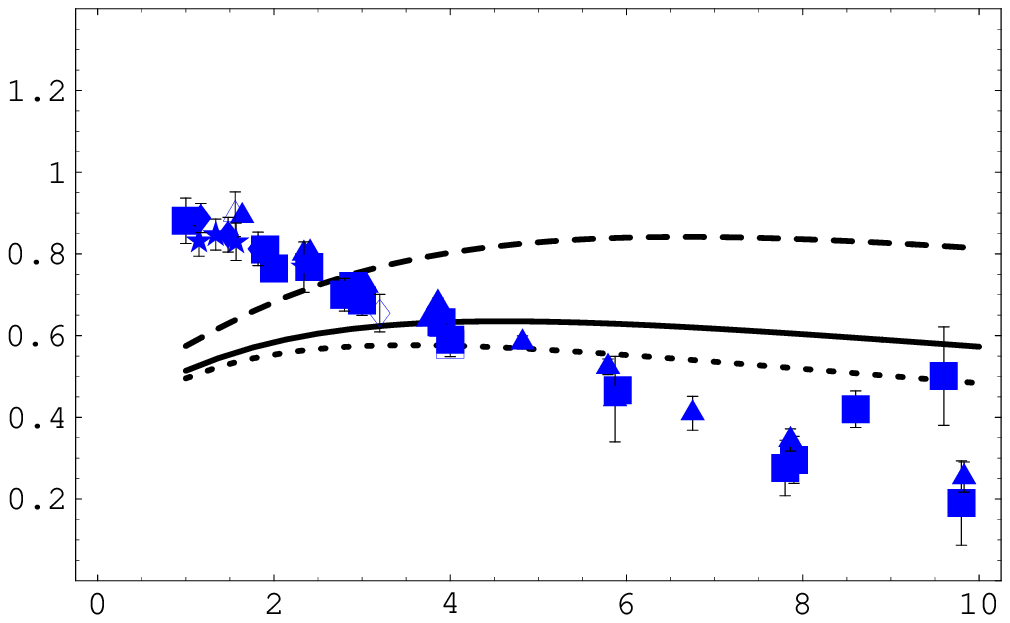}
  \includegraphics[width=0.33\textwidth,angle=0]{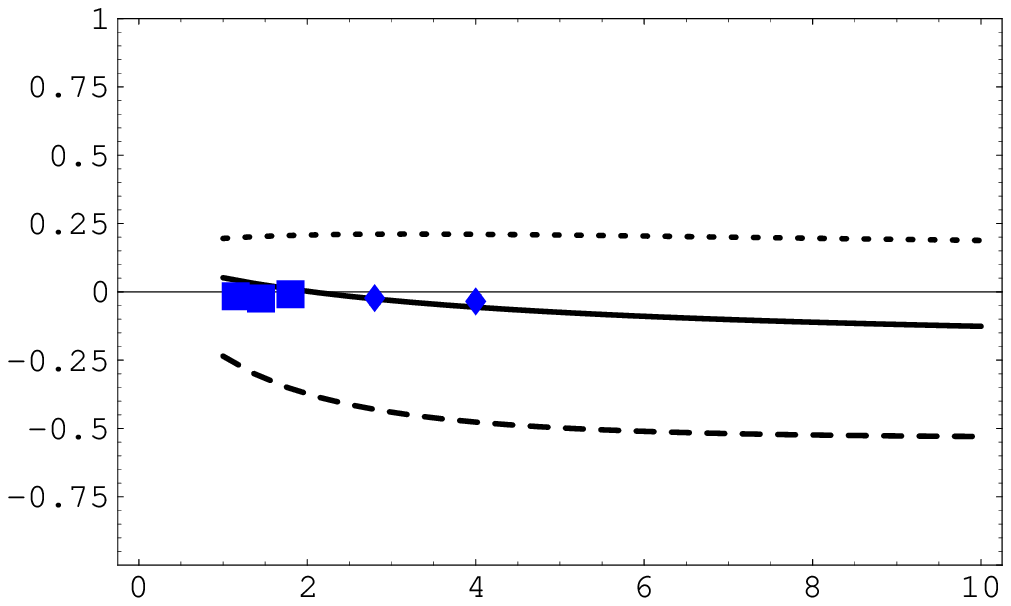}
  \includegraphics[width=0.33\textwidth,angle=0]{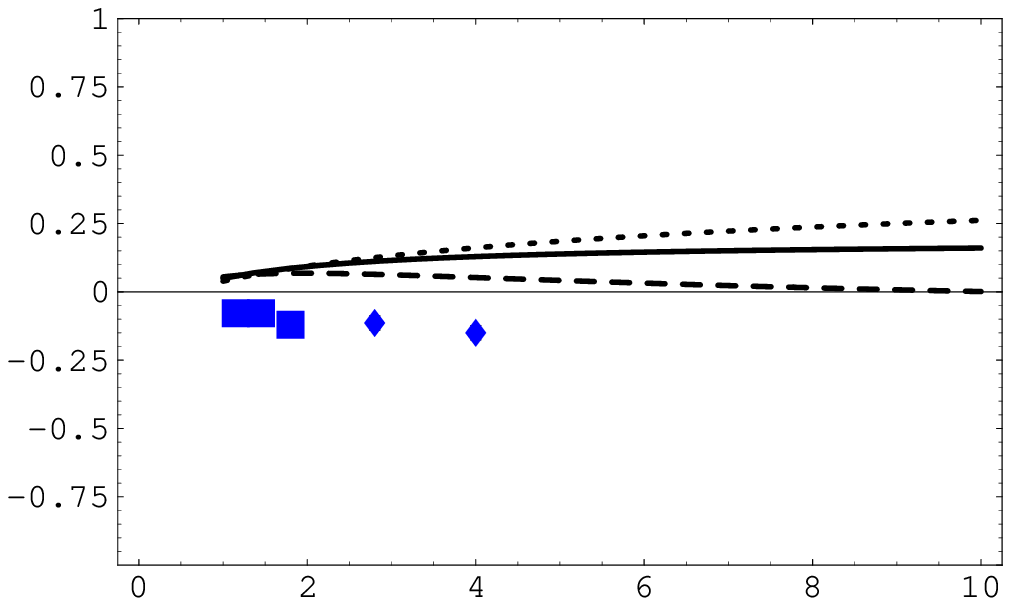}
\caption{$\gamma* N \to \Delta$ transition form factors 
(left:$G_M^*/(3 G_{\it Dipole})$ vs. $Q^2$, middle:$R_{EM}$  vs. $Q^2$, right: $R_{SM}$  vs. $Q^2$)
in  the LCSR approach~\cite{BLPR2005}
obtained using the BLW model (solid line), 
the asymptotic model (dashed line)
and sum rule model (dotted line) of the nucleon DAs.
For the references to the experimental data see ~\cite{BLPR2005}.
(Color identification refers to the online version)}
\label{fig:Delta}
\end{figure*}

The nucleon DAs provide the principal nonperturbative input to the LCSRs. 
We use here three models for the nucleon distribution amplitude:
the {\it asymptotic} form (dashed lines), the {\it (QCD) sum-rule} estimate (dotted
lines) and the
{\it BLW} model (solid lines).
The corresponding numerical values can be found in \cite{BLW2006}.
To our accuracy, the sum rules for the nucleon form factors do not depend on the parameters 
$\lambda_2$ and $f_2^d$;
this dependence is present, however, in the transition form factors of 
$\gamma^* N \to \Delta$.    
\begin{figure}[ht]
  \includegraphics[width=0.45\textwidth,angle=0]{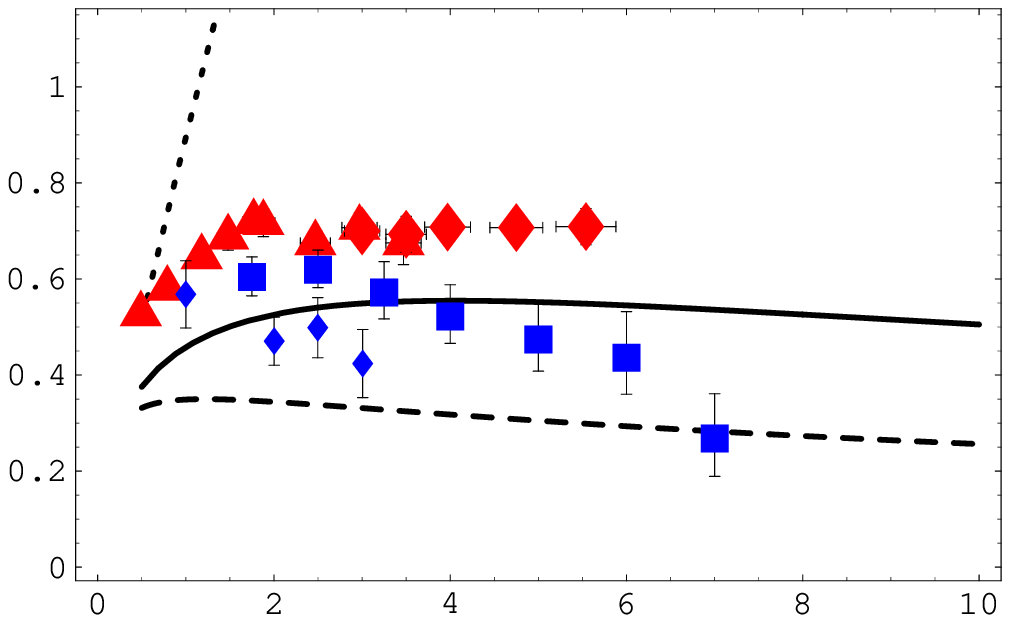}
\caption{LCSR results (solid curves) for the ratio  $\sqrt{Q^2} F_2^p/( F_1^p 1.79)$ 
obtained using the BLW model (solid line), 
the asymptotic model (dashed line)
and sum rule model (dotted line) of the nucleon DAs.
{\it Red symbols}: experimental values obtained via Polarization transfer:  
{\it Blue symbols}: experimental values obtained via Rosenbluth
separation.
For the references to the experimental data see ~\cite{BLW2006}.
} 
\label{fig:F2/F1}
\end{figure}
One sees that the experimental data on the electromagnetic form factors are reproduced very well, 
and, most welcome, the unphysical tensor form factor $G_T$ is consistent with zero. 
Also for the axial form factor there is
a good agreement, both in shape and normalization.
In particular for $G_E^p / G_M^p$ we have a very strong dependence on the form of the
nucleon distribution amplitude.
We also calculated the
$\gamma^* N \to \Delta$ transition form factors within the LCSR approach, see \cite{BLPR2005}.
The results are shown in Fig.~\ref{fig:Delta}. In this case we also get a relatively good
agreement with the experimental data. 
\\   
We should warn that the BLW model from \cite{BLW2006} is not based on any systematic attempt to fit
the data and in fact we believe that any fitting would be premature before the radiative 
corrections to the LCSR are calculated. 
In addition, one has to take into account the 
scale dependence of the parameters of the DAs and study in more detail the dependence 
of the sum rules on the Borel parameter.  
Still, the very
possibility to describe many different form factors using the same set of DAs is nontrivial
and indicates the selfconsistency of our approach. 

\section{Outlook}
In order to make our approach really quantitative one should also
include $\alpha_s$-corrections to the light-cone sum rules for the 
form factor. This work is currently in progress.
Having the $\alpha_s$-corrections at hand, one can fit
the light-cone sum rules to the experimental values and therefore determine
the nucleon distribution amplitude.
\\
Moreover it would also be desirable to have some lattice determination of all
non-perturbative parameters of the nucleon distribution amplitudes.
\\
Finally from a phenomenological point of view it might be very interesting
to apply this formalism to $\Lambda_b$ decays. First steps in that direction have 
already been performed in \cite{HW2004,HW2006}.
There are currently some data from TeVatron and there will be much more 
from the LHC for the heavy baryons.
Investigating the $\Lambda_b$ baryon is in particular interesting since there
seems to be some problem with the lifetimes. Experimentally one always obtained
values for $\tau(\Lambda_b) / \tau (B_d)$ smaller than one \cite{PDB}.
Inclusive decays of heavy hadrons, see e.g. \cite{LNO1997,LNO1998,L2000}
can be calculated within the framework of the Heavy Quark Expansion
(HQE), e.g. \cite{HQE} as an expansion in inverse power of the heavy $b$-quark mass.
In that approach the dominant contributions to the lifetime ratios arise 
typically at the third order of the HQE,
therefore the calculation is analogous to the mixing quantities in the neutral $B$-system, see e.g.
\cite{BBGLN98,L1999,BL2000,BBLN2003,L2004,LN2006,L2007b}.
The lifetime ratios of heavy hadrons have been theoretically considered e.g. in
\cite{L2001,LW2001,BBGLN2002,Cern2003,L2006,L2007,Cecilia} and one obtains typically values for 
$\tau(\Lambda_b) / \tau (B_d)$ closer to one. 
Before speaking from a real discrepancy one has to keep in mind however, that 
$\tau(\Lambda_b) / \tau (B_d)$  is theoretically in a much worse shape than
$\tau(B^+) / \tau (B_d)$, see e.g. \cite{L2007}.
Moreover there are now new measurements of $\tau(\Lambda_b)$ \cite{CDFLambdab,D0Lambdab}
in non-leptonic channels 
on the market, that do not agree with each other. So the $\Lambda_b$-system
is awaiting some theoretical and experimental progress.

\begin{theacknowledgments}
I would like to thank V. Braun, D. Ivanov, N. Mahnke, 
A. Peters, G. Peters, A. Radyushkin, E. Stein and M. Wittmann for the 
pleasant collaboration and the organizers of this workshop for their perfect 
work.

\end{theacknowledgments}

\section{Appendix}
Unfortunately there were was a sign error in the $x^2$-correction
${A}_1^{M(u)}$ in appendix C of \cite{BLW2006}.
Eqs. (C23-C25) in \cite{BLW2006} have to be replaced by

\begin{eqnarray}
{\cal T}_1^{M(u)} (x) & = &  \frac{1}{2} 
\left[ V_1^{M(d)} (x) + V_1^{M(u)} (x) +  A_1^{M(u)} (x) \right],
\nonumber \\
{\cal T}_1^{M(d)} (x) & = &  V_1^{M(u)} (x) -  A_1^{M(u)} (x) \,.
\end{eqnarray}

\begin{eqnarray}
{\cal T}_1^{M(u)} (x)  & = & \frac{ x^2}{48}  \left( f_N E_{f}^u  + \lambda_1 E_{\lambda}^u  \right),
\nonumber \\
{\cal T}_1^{M(d)} (x)  & = & \frac{ x^2 (1\!-\!x)^4}{4}  \left( f_N E_{f}^d  + \lambda_1 E_{\lambda}^d  \right)
\end{eqnarray}
with
\begin{eqnarray}
E_{f}^u & = & -\Big[
(1 - x)
\left(
3 (439 + 71 x - 621 x^2 + 587 x^3 - 184 x^4)
\right.
\nonumber\\&&{}
+ 4 A^u_1 (1 - x)^2 (59 - 483 x + 414 x^2)
\left.
- 4 V^d_1 (1301 - 619 x - 769 x^2 + 1161 x^3 - 414 x^4)
\right)
\Big] 
\nonumber\\&&{}
- 12(73 - 220V_1^d) \ln[x]\,,
\nonumber \\
E_{\lambda}^u & = & -\Big[
(1 - x) 
( 5 - 211 x + 281 x^2 - 111 x^3
\nonumber\\&&{} +10 (1 +  61 x -  83 x^2 +  33 x^3) f_1^d - 40(1 - x)^2 (2 - 3 x) f_1^u)
\Big] 
- 12 (3 - 10 f_1^d) \ln[x]\,,
\nonumber \\
E_{f}^d & = & 17 + 92 x + 12 (A^u_1+V^d_1) (3 - 23 x) 
\,,
\nonumber \\
E_{\lambda}^d & = & -7 + 20 f^d_1 + 10 f^u_1
\,. 
\end{eqnarray}
Note that the $x^2$-corrections do not depend on $\lambda_2$ and $f^d_2$.

\bibliography{Lenz}

\end{document}